\def\etal{{\it et al.\thinspace}}
\def\eg{{\it e.g.\ }}

\def\ie{{\it i.e.\ }}

\def\gsim{~\rlap{$>$}{\lower 1.0ex\hbox{$\sim$}}}
\def\lsim{~\rlap{$<$}{\lower 1.0ex\hbox{$\sim$}}}

\documentclass[12pt,preprint]{aastex}
\begin{document}

\title{Apsidal Behavior Among Planetary Orbits: Testing the Planet-Planet Scattering Model}

\author{Rory Barnes\altaffilmark{1} and Richard Greenberg\altaffilmark{1}}

\altaffiltext{1}{Lunar and Planetary Laboratory, University of Arizona,
Tucson, AZ 85721, rory@lpl.arizona.edu, greenber@lpl.arizona.edu}


\begin{abstract}
Planets in extrasolar systems tend to interact such that their orbits
lie near a boundary between apsidal libration and circulation, a
``separatrix'', with one eccentricity periodically reaching
near-zero. One explanation, applied to the $\upsilon$ And system,
assumed three original planets on circular orbits. One is ejected,
leaving the other two with near-separatrix behavior. We test that
model by integrating hundreds of hypothetical, unstable planetary
systems that eject a planet. We find that the probability that the
remaining planets exhibit near-separatrix motion is small ($< 5\%$
compared with nearly 50\% of observed systems). Moreover, while observed
librating systems are evenly divided between aligned and anti-aligned
pericenter longitudes, the scattering model strongly favors
alignment. Alternative scattering theories are proposed, which may
provide a more satisfactory fit with observed systems.
\end{abstract}

\section{Introduction}

A significant fraction ($\sim 50$\%) of adjacent pairs of planets lie
near a ``secular separatrix'', \ie a boundary in orbital element space
between apsidal libration (the difference in the longitudes of
periastron $\Delta\varpi$ oscillates about a fixed value) and
circulation ($\Delta\varpi$ oscillates through 360$^o$) (Barnes \&
Greenberg 2006a, 2006b). One characteristic of this behavior is that
one orbit periodically becomes circular. For the $\upsilon$ And
system, an archetypal near-separatrix system, Ford \etal (2005)
suggested a model involving an unstable system of three planets on
initially coplanar, circular orbits in which a gravitational encounter
ejects one planet, leaving a pair of planets still bound to the
star. The event created a new ``initial condition'' with one planet on
an eccentric orbit and the other still on a nearly circular orbit,
such that the subsequent secular interaction is near-separatrix.

This ``planet-planet scattering'' model was first posited to explain
the large eccentricities of extra-solar planets (Rasio \& Ford
1996). Malhotra (2002) invoked a simplified version (one planet was
massless) of this model to explain what was then believed to be a high
fraction of systems exhibiting apsidal libration (\eg Zhou \& Sun
2003).

In fact, based on improved observations and statistics, Barnes \&
Greenberg (2006a) found that libration is relatively rare ($\sim$15\% of cases). What is common, whether a system librates or circulates,
is to be near the boundary between those states. Ford
\etal (2005) described only one specific hypothetical case that, when
integrated numerically, did result in two planets near a secular
separatrix. However, that case is anecdotal. Here we consider whether
simulations like those in the Ford \etal model of planet-planet
scattering can statistically reproduce the observed large fraction of
systems that lie near the secular separatrix, as well as the observed
distribution among circulating, aligned librating and anti-aligned
librating systems (which Ford \etal does not address). Here we
systematically survey hundreds of initial conditions, similar to the
case described in Ford \etal, in order to consider whether
planet-planet scattering can explain the characteristics of the
observed systems.

\section{Methodology}

We consider a hypothetical system of a 1.3 M$_\odot$ star and three 
planets, called 1, 2, and 3, with respective semi-major axes of 0.83, 
3.555 and 4.4 AU, and masses of 1.94, 3.94 and 1.32 Jupiter masses 
(properties similar to the observed $\upsilon$ And system and the 
hypothetical configuration considered in Ford \etal). All these orbits are 
circular and coplanar. We choose an initial condition with planet 1 45$^o$ 
ahead of planet 3 in longitude, $L$. We then consider 360 similar cases, 
but with the initial longitude of planet 2, $L_2$, distributed evenly 
around 360$^o$ in 1$^o$ intervals. With these masses and orbits, the outer 
two planets fail a known stability condition (Gladman 1993; Barnes \& 
Greenberg 2006c), which is independent of $L$. We also considered a 
sampling of cases with different semi-major axes and masses.

We use the symplectic, $N$-body integrator MERCURY (Chambers 1999) to
integrate  each case for $10^5$
years. We require each simulation to conserve energy to within 1 part in
$10^4$, which has been shown to be sufficiently accurate for
symplectic integration methods (Barnes \& Quinn 2004). Our smallest
timestep was $10^{-3}$ days. For this level of energy conservation, angular momentum conservation was always at least 1 order of magnitude better. For
configurations that eject the outer planet, we then
integrate the remaining planets for $10^5$ years in order to characterize the secular interaction of those remaining planets, \ie to calculate
the orbit's proximity to the separatrix and to determine the type of
apsidal interaction. In a few cases we integrated for $5\times 10^5$
years because the resultant secular period was longer than $10^5$
years.

\section{Results}
After 100,000 years, of the sample of 360 cases with varying $L_2$, 169 ejected the outer
planet leaving two planets engaged in ongoing secular
interactions that could be characterized in meaningful
ways. An additional 95 cases also ejected the outer planet,
but left one of the remaining planets with $a > 6$ AU, too far to be
observed by current search methods. Of the remaining cases,
37 ejected no planets, but left all of them on highly eccentric orbits
that interacted in complex and chaotic ways. Another 5
ejected only the middle planet, again leaving planets in highly
eccentric, unstable, chaotically interacting orbits. In 49
cases two planets were ejected. Finally, 5 cases were
rejected on the technical grounds that energy was not conserved to a
level that could guarantee our desired precision. This
number of cases is too small to affect the resulting statistics.

In order to have an outcome with near-separatrix motion, as envisioned
for example in the scenario proposed by Ford \etal, a certain
sequence of events appears to be required. Fig.\ 1a shows the evolution
of one of our cases that does yield such an outcome. First,
within only a few years, an interaction between planets 2 and 3 yields $e_3 > 0.7$ and a substantial increase in $e_2$. Planet 3 spends most of its time far from the other planets, so the
inner two planets undergo secular interactions independent of the
third. Because $e_2$ becomes non-zero while the inner orbit
remains circular, their secular behavior is characterized by periodic
returns of $e_1$ to zero and typical near-separatrix behavior.
Eventually, the outer planet might have a close encounter with one or
both of the inner planets, which would wreak havoc with the regular
secular behavior. However, before that can happen, within a
few thousand years, planet 3 receives a small kick that ejects it from
the system. The kick required to eject planet 3 is small enough that it does not significantly affect the secular interaction of 1 and 2, and their near-separatrix behavior is preserved.

Thus the requirements for near-secular behavior seem to be (1) a quick 
large
increase in $e_3$, with a modest increase in $e_2$ while $e_1$ remains zero,
followed by (2) an encounter that ejects planet 3 without disturbing
the other planets too much.

\clearpage
\begin{figure}
\plotone{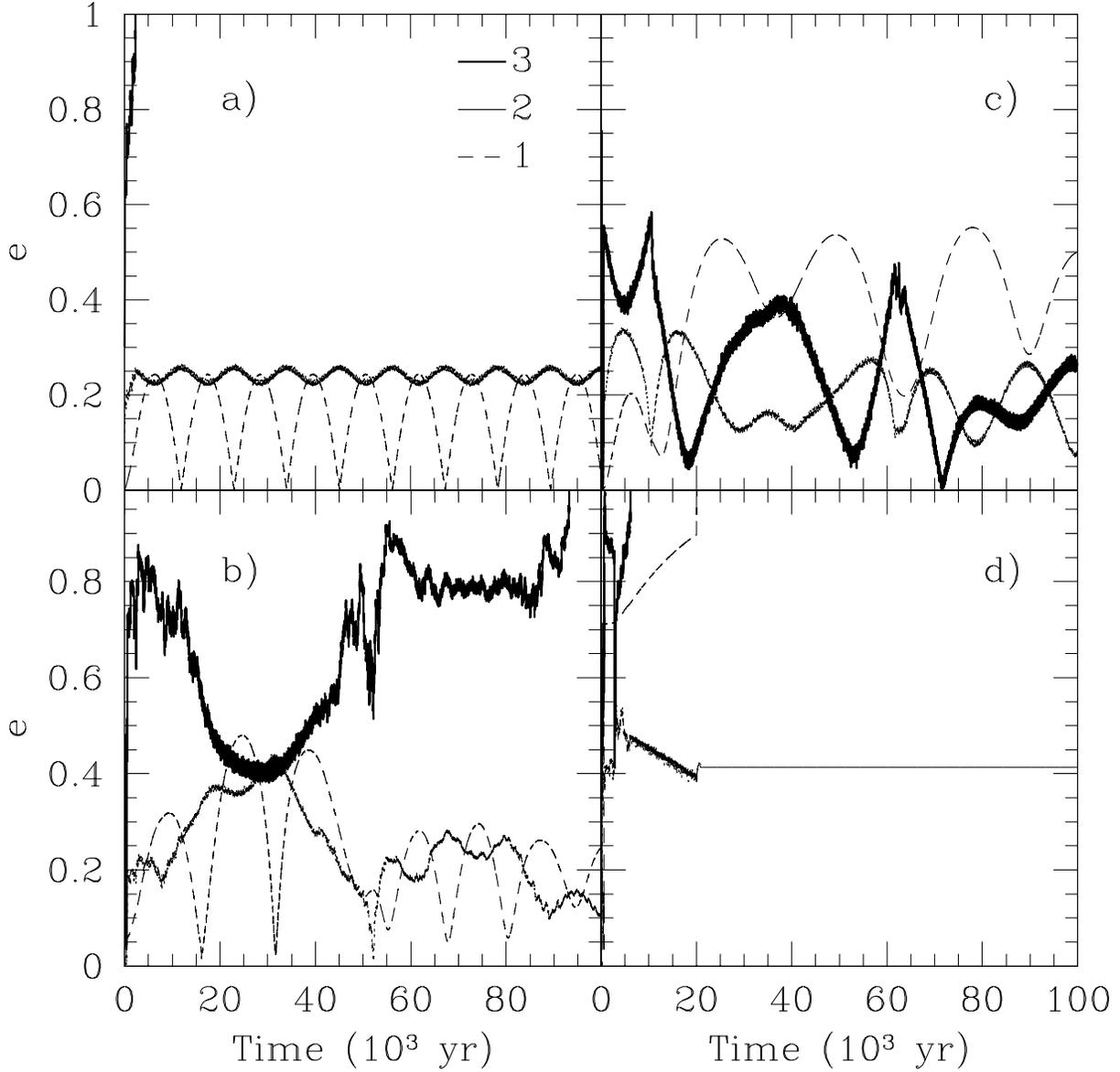}
\figcaption[]{\label{fig:com4} \small{Evolution of orbital eccentricities in four
characteristic cases. (a) This case, with the initial longitude of
planet 2 relative to planet 3 of 123$^o$, shows a typical series of
events leading to near-separatrix motion. (b) In this case, the initial
encounter starts near-separatrix behavior of planets 1 and 2, but planet 3
remains in the system and disrupts the regular secular behavior $L_2 =
274^o$. (c) Here, the initial encounter fails to increase $e_3$ enough to
uncouple it from the other planets ($L_2 = 90^o$). (d) An adequate early
jump in $e_3$, is soon followed by disruptive interactions that eject planet 3. However the remaining
planets' secular interaction ultimately results in a destabilizing encounter ($L_2 = 314^o$).}}
\end{figure}
\clearpage

Fig.\ 1b shows a case that satisfies the first condition, but
not the second. Planet 3 quickly enters a highly eccentric orbit, and
the other planets begin to behave like a near-separatrix case, but planet
3 isn't immediately ejected. By 10,000 years it begins to disrupt the regular
two-planet secular behavior of the other planets, allowing them to evolve
onto orbits where $e_1$ never returns to zero, \ie no longer near a
separatrix.

Fig.\ 1c shows what can happen if the first requirement is not
met. Here the initial increase in $e_3$ is too little to keep it out of
the way of the other planets. This example is one of the 37 cases that
left all three planets interacting in ways that preclude regular
secular behavior.

Fig.\ 1d shows a case where planet 3 does satisfy the first
requirement, and it is also ejected from the system fairly quickly (in
$\sim$10,000 yr as in Fig. 1a), but before being ejected it has
encountered and grossly de-circularized the other two orbits. After planet 3 is
ejected, planets 1 and 2 begin a secular interaction, but this interaction leads to large values of $e_1$ and an encounter that ejects planet
1. 

Next we characterize the outcomes of our cases for comparison with Ford
\etal 's suggestion that this process can explain behavior like that of
the near-separatrix $\upsilon$ And system, and also for comparison
with the more general statistics of observed systems. A way to
quantify how close a given system is to a separatrix was introduced by
Barnes and Greenberg (2006a). Loosely described, a parameter
$\epsilon$ represents the ratio between the minimum $e$ value and the
amplitude of oscillation of $e$ (see Barnes and Greenberg [2006a] for
a precise definition). Small $\epsilon$ means the system is
near-separatrix.

Figure 2 shows the distribution of $\epsilon$ values as a fraction of
the 360 initial cases. Only the 169 cases that produced regular
secular behavior contribute to these statistics; for the remainder of
the cases $\epsilon$ would be meaningless and near-separatrix motion
is out of the question. The distribution shows a very slight rise for
small values of $\epsilon$: About 4\% have $\epsilon$ smaller than
0.01 and 12\% have values less than 0.03. We also show the
distribution for the subset of cases in which the outer planet was
ejected in only 20,000 years, with a similar distribution. 

All simulations that resulted in $\epsilon < 0.01$ had final $a_2$
(semi-major axis of planet 2) in the range $2.84 \pm 0.04$
AU. Overall, 30\% of simulations ended with $a_2$ in this range.

In addition to varying the initial longitude, we performed 8 integrations in which $a_2$
was varied by 0.01 AU, and 8 which varied the mass of planet 2 by 0.01
Jupiter masses. Planet-planet scattering in these cases also resulted
in 1 configuration near the secular separatrix, consistent with Fig.\ 2.

\clearpage
\begin{figure}
\plotone{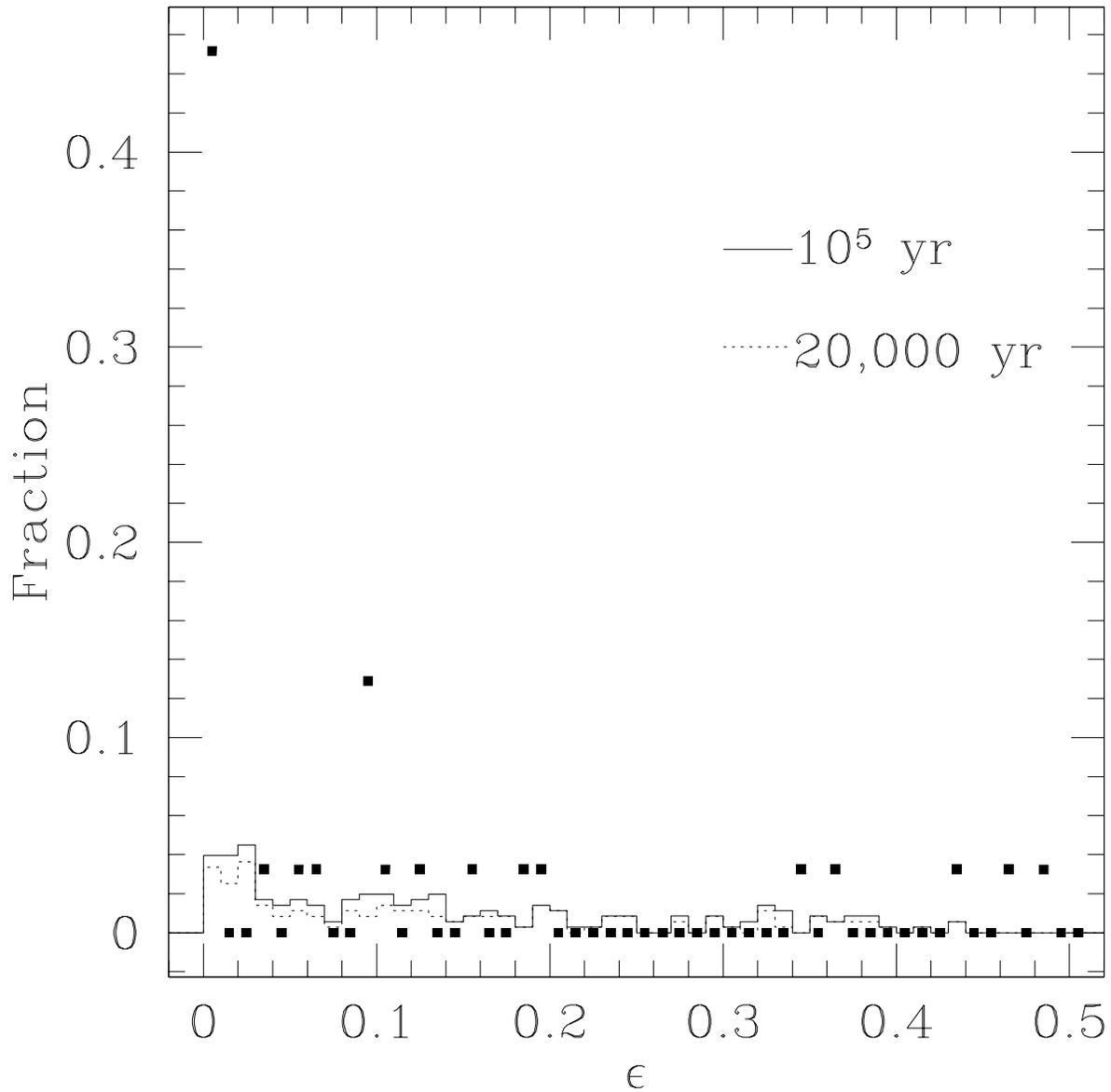}
\figcaption[]{\label{fig:epscomp} \small{Distribution of $\epsilon$, a parameter which describes how
close a system is to a separatrix. Squares represent the distribution
for observed systems. The solid line shows statistics for our calculated
outcomes of the planet-planet scattering model proposed by Ford \etal
(2005). The real systems tend to be much closer to the separatrix (small
$\epsilon$) than can be reproduced by the Ford \etal hypothesis. The
dashed line shows the subset of the modeled cases where ejection of
planet 3 occurred within 20,000 yr. The bin size for these histograms is
0.01.}}
\end{figure}
\clearpage

The distribution of $\epsilon$ values among actual observed planetary
systems is much more concentrated near the separatrix, with very small
$\epsilon$ values, as reported by Barnes and Greenberg (2006a).  In
Fig. 2 we include the statistics for the observed systems for
comparison with the results generated by the planet-planet scattering
model. For the observed values of $\epsilon$, we use the results from
Barnes and Greenberg (2006a), plus values of $\epsilon$ calculated for
the two recently discovered systems HIP 14810 (Wright \etal 2007) and
HD 160691 (Pepe \etal 2007), see the Appendix. Note that $e$ is poorly
known for 47 UMa and GJ 876 (Butler \etal 2006), but the statistics
are not affected by these two systems. Similarly, the result would be
unchanged if we excluded resonant and/or tidally-evolved systems
(Barnes \& Greenberg 2006a). As shown in Fig 2, in the observed
distribution, 45\% of the systems have $\epsilon$ smaller than
0.01. (In fact, the results for HIP 14810 and HD 160691 in the
Appendix show the same distribution; two of the four pairs evolve with 
$\epsilon < 0.01$.)

Based on our numerical experiments, the planet-planet scattering model
as described by Ford \etal (2005) does not seem to reproduce the
observed strong tendency for systems to lie near a separatrix. While
selected orbital configurations can lead to a small-$\epsilon$ system
like $\upsilon$ And, in general this model does not appear to reproduce
that large fraction of systems that exhibit behavior very near a
secular separatrix. 

Of course, our experiments sample only a small portion of the possible 
multi-dimensional parameter space, but it gives a more reliable estimate 
of the probability of outcomes than the single case shown in Ford \etal 
Moreover, our tests show one specific sequence of events (\eg Fig.\ 1a) 
leads to near-separatrix motion in the manner envisioned by Ford \etal It 
would be surprising if such a sequence were common, given the results 
presented here. Other sequences may be possible, but they are not present 
in our simulations. Further tests would be useful, but it seems unlikely 
that the planet-planet scattering model of Ford \etal can explain the 
preponderance of near-separatrix systems.

Another disagreement between the statistical outcomes of the
planet-planet scattering model and actual systems comes from
consideration of the modes of libration. In secular interactions,
orbits can librate about an alignment of their major axes with their
pericenters at the same longitude or an anti-alignment where the
pericenters are 180$^o$ apart. Among observed systems (again based on
calculations from Barnes and Greenberg [2006a] and the Appendix
herein), there is nearly equal division between systems librating
about aligned pericenters (2 out of 31) and those librating about
anti-aligned pericenters (also 2 out of 31), with the remaining 87\%
circulating rather than librating. Among the systems generated by the
planet-planet scattering model, librating cases overwhelmingly favor
aligned pericenters (by a factor $>10$) relative to
anti-alignment. Specifically, 24\% are aligned and less than 2\% are
anti-aligned. The planet-planet scattering model does not produce
systems consistent with the population of observed planets.

\section{Conclusions}
Our numerical experiments demonstrate that the planet-planet
scattering process, as described by Ford \etal (2005), does not
explain the prevalence of near-separatrix apsidal behavior, even
though the model had been proposed to explain this type of motion in
the case of $\upsilon$ And. That model can yield near-separatrix
behavior, but the probability is too low for it be a significant
factor. The model does produce systems consistent with real systems in
one respect: Most of them undergo apsidal circulation rather than
libration.  However, it produces about twice as many librating cases
as are observed in reality, and it yields far too small a portion of
those in anti-aligned libration compared with observed systems.

Another problem with Ford \etal 's planet-planet scattering is that
the initial set up involves planets on circular orbits that are too
close to be stable.  It is difficult to envision a formation process
that could yield such a system.  For example, if a hypothetical
eccentricity-damping medium were invoked to explain the circular
orbits, the medium would need to disappear in less than a synodic
period to provide the initial condition in the planet-planet
scattering scenario. Even if an explanation can be found for such an
initial set-up, our experiments suggest that the statistics of
observed behavior would not follow.

We propose a modification to the scattering model that
may explain the propensity for producing near-separatrix orbits based
on conditions consistent with other lines of evidence about the
origins of planetary systems.  From our simulations of scattering in
this current study, we find that the key features of those scattering
events that lead to near-separatrix behavior are an abrupt, modest
increase in the eccentricity of one planet, while another planet
remains on a circular orbit, followed by a rapid removal of the cause
of the perturbation (a third planet) from the system.  The problems
with starting the perturbing planet on a circular orbit are (1) that
it is hard to understand why an unstable orbit would be circular (as
mentioned above) and (2) after the initial encounter, this planet is
rarely ejected from the system soon enough to keep from further
modifying the interaction of the other planets.

Suppose instead that the perturbing planet started not on an unstable
circular orbit, but rather on a long-period, high-eccentricity
orbit. Such high-eccentricity bodies scattered about the solar system
have been invoked to explain basic properties of the system, including
the origin of the Moon by an impact into the Earth, pumping the
relative velocities among asteroids to suppress planet growth in that
zone, and generating the late heavy bombardment (Gomes \etal 2005;
Tsiganis \etal 2005). Indeed the temporary passage of large bodies
scattered from the outer solar system, at a time when the inner
planets had achieved fairly stable, near-circular orbits is a standard
component of current models of the formation of our solar system (\eg
Gomes \etal 2005; Strom \etal 2006). We call this model the Rogue Planet 
Model.

As such a rogue planet or protoplanet passed through the inner part of a
planetary system, eventually it would pass close enough to one of the
regular (circularly orbiting) planets to impose a velocity change and
introduce some orbital eccentricity. At the same time, because the
impulsive perturber was on an extended orbit from the outer part of the
system, there would be a substantial probability that it would be ejected
from the system by the same encounter, preventing any further impulses on
inner-system planets. After the encounter, any other regular planet
will still be on a near-circular orbit, so that the subsequent secular
behavior will be near-separatrix.

In order to test our proposed modification of the planet scattering
model, the hypothesis should be explored with numerical experiments,
analogous to those presented here and in Gomes \etal (2005).  In our
hypothesis, the prevalence of near-separatrix behavior in extra-solar
planetary systems, as well as many of the dynamical properties of our Solar
System, can be described by late-stage scattering of protoplanets. If
these bodies are scattered inward, toward planets on circular, stable
orbits, the protoplanets are ejected, leaving planets on near-separatrix
orbits. If scattered out, they may become the cores of ice giants, or 
become part of a scattered disk
component of a Kuiper Belt. If this hypothesis is correct, then the
origin of the Solar System's small eccentricities, extra-solar
planets' large eccentricities, and all planetary systems' tendency to
lie near a secular separatrix is explained by a single model.

\smallskip
This work was funded by NASA's Planetary Geology and Geophysics
program grant NNG05GH65G. We thank Fred Rasio, Eric Ford and an anonymous referee for useful suggestions that clarified this manuscript.

\references
Barnes, R. \& Greenberg, R. 2006a, ApJ, 652, L53\\
------------. 2006b, ApJ, 638, 478\\
------------. 2006c, ApJ, 647, L163\\
Barnes, R. \& Quinn, T.R. 2004, ApJ, 611, 494\\
Butler, R.P. \etal 2006, ApJ, 646, 505\\
Chambers, J., 1999, MNRAS, 304, 793\\
Ford, E.B., Lystad, V., \& Rasio, F.A. 2005, Nature, 434, 873\\
Gladman, B. 1993, Icarus, 106, 247\\
Gomes, R. \etal 2005, Nature, 435, 466\\
Malhotra, R. 2002, ApJ, 575, L33\\
Pepe, F. \etal 2007, A\&A, 462, 769\\
Rasio, F.A., \& Ford, E.B. 1996, Science, 274, 954\\
Strom, R.G. \etal 2006, Science, 309, 1847\\
Tsiganis \etal 2005, Nature, 435, 459\\
Wright, J.T. \etal 2007, ApJ in press\\
Zhou, J.-L., \& Sun, Y.-S. 2003, ApJ, 598, 1290\\

\section*{Appendix}
Since the publication of Barnes \& Greenberg (2006a), two planetary
systems have been announced or revised. HIP 14810 (Wright \etal 2007)
has two planetary mass companions, one of which has been tidally
circularized. HD 160691, also called $\mu$ Ara, now has four planets
(Pepe \etal 2007), of which the innermost is also tidally evolved. HD 160691 is
unstable over long timescales ($\sim 10^8$ years), and its properties
are therefore especially suspect. We have performed a dynamical
analysis of the best-fit orbits to these two systems in the same
manner as Barnes \& Greenberg (2006a) in order to calculate the
apsidal behavior and $\epsilon$. These properties are listed in Table
1, where C stands for circulation and L$_{180}$ for anti-aligned
libration. The designation C/C, as in Barnes \& Greenberg (2006a), means the system
lies near a ``circulation mode separatrix''.

\medskip
\begin{center}Table 1: Apsidal Properties of HIP 14810 and HD 160691\\
\begin{tabular}{cccc}
\hline
System & Pair & $\epsilon$ & Apsidal Behavior\\
\hline\hline
HIP 14810 & b-c & 0.05 & L$_{180}$\\
HD 160691 & c-d & 0.002 & C/C\\
 & d-b & 0.003 & C/C\\
 & b-e & 0.13 & C\\
\end{tabular}
\end{center}

\end{document}